\documentclass[superscriptaddress,aps,amsmath,amssymb,floatfix,reprint,raggedbottom,longbibliography]{revtex4-1}
\usepackage{graphicx}
\usepackage[colorlinks=true,citecolor=blue,linkcolor=blue,urlcolor=blue]{hyperref}
\usepackage{dcolumn}
\usepackage{xcolor}
\usepackage{fancyhdr}
\usepackage{lipsum}
\usepackage{soul}
\usepackage{braket}

\makeatletter 
\renewcommand{\fnum@figure}{\textbf{FIG.~\thefigure}}
\makeatother

\makeatletter
\def\bbordermatrix#1{\begingroup \m@th
  \@tempdima 4.75\p@
  \setbox\z@\vbox{%
    \def\cr{\crcr\noalign{\kern2\p@\global\let\cr\endline}}%
    \ialign{$##$\hfil\kern2\p@\kern\@tempdima&\thinspace\hfil$##$\hfil
      &&\quad\hfil$##$\hfil\crcr
      \omit\strut\hfil\crcr\noalign{\kern-\baselineskip}%
      #1\crcr\omit\strut\cr}}%
  \setbox\tw@\vbox{\unvcopy\z@\global\setbox\@ne\lastbox}%
  \setbox\tw@\hbox{\unhbox\@ne\unskip\global\setbox\@ne\lastbox}%
  \setbox\tw@\hbox{$\kern\wd\@ne\kern-\@tempdima\left[\kern-\wd\@ne
    \global\setbox\@ne\vbox{\box\@ne\kern2\p@}%
    \vcenter{\kern-\ht\@ne\unvbox\z@\kern-\baselineskip}\,\right]$}%
  \null\;\vbox{\kern\ht\@ne\box\tw@}\endgroup}
\makeatother

\begin{document}

\title{Scalable Emulation of Sign-Problem$-$Free Hamiltonians\\ with Room Temperature p-bits}
\author{Kerem Y. Camsari, Shuvro Chowdhury and Supriyo Datta}
\affiliation{School of Electrical and Computer Engineering, Purdue University, IN, 47907}
\date{\today}
\begin{abstract}
The growing field of quantum computing is based on the concept of a q-bit which is a delicate superposition of 0 and 1, requiring cryogenic temperatures for its physical realization along with challenging coherent coupling techniques for entangling them. By contrast, a probabilistic bit or a p-bit is a robust classical entity that fluctuates between 0 and 1, and can be implemented at room temperature using present-day technology. Here, we show that a probabilistic coprocessor built out of room temperature p-bits can be used to accelerate simulations of a special class of quantum many-body systems that are sign-problem$-$free or  ``stoquastic'',  leveraging the well-known Suzuki-Trotter decomposition that maps a $d$-dimensional quantum many body Hamiltonian to a $d$+1-dimensional classical Hamiltonian. This mapping allows an efficient emulation of a quantum system by classical computers and is commonly used in software to perform Quantum Monte Carlo (QMC) algorithms. By contrast, we show that a compact, embedded MTJ-based coprocessor can serve as a highly efficient hardware-accelerator for such QMC algorithms providing several orders of magnitude improvement in speed compared to optimized CPU implementations. Using realistic device-level SPICE simulations we demonstrate that the correct quantum correlations can be obtained using a classical p-circuit built with existing technology and operating at room temperature. The proposed coprocessor can serve as a tool to study stoquastic quantum many-body systems, overcoming challenges associated with physical quantum annealers.

  \end{abstract}
\maketitle

\section{Introduction}

\begin{figure*}[t!]
\includegraphics[width=0.85\linewidth]{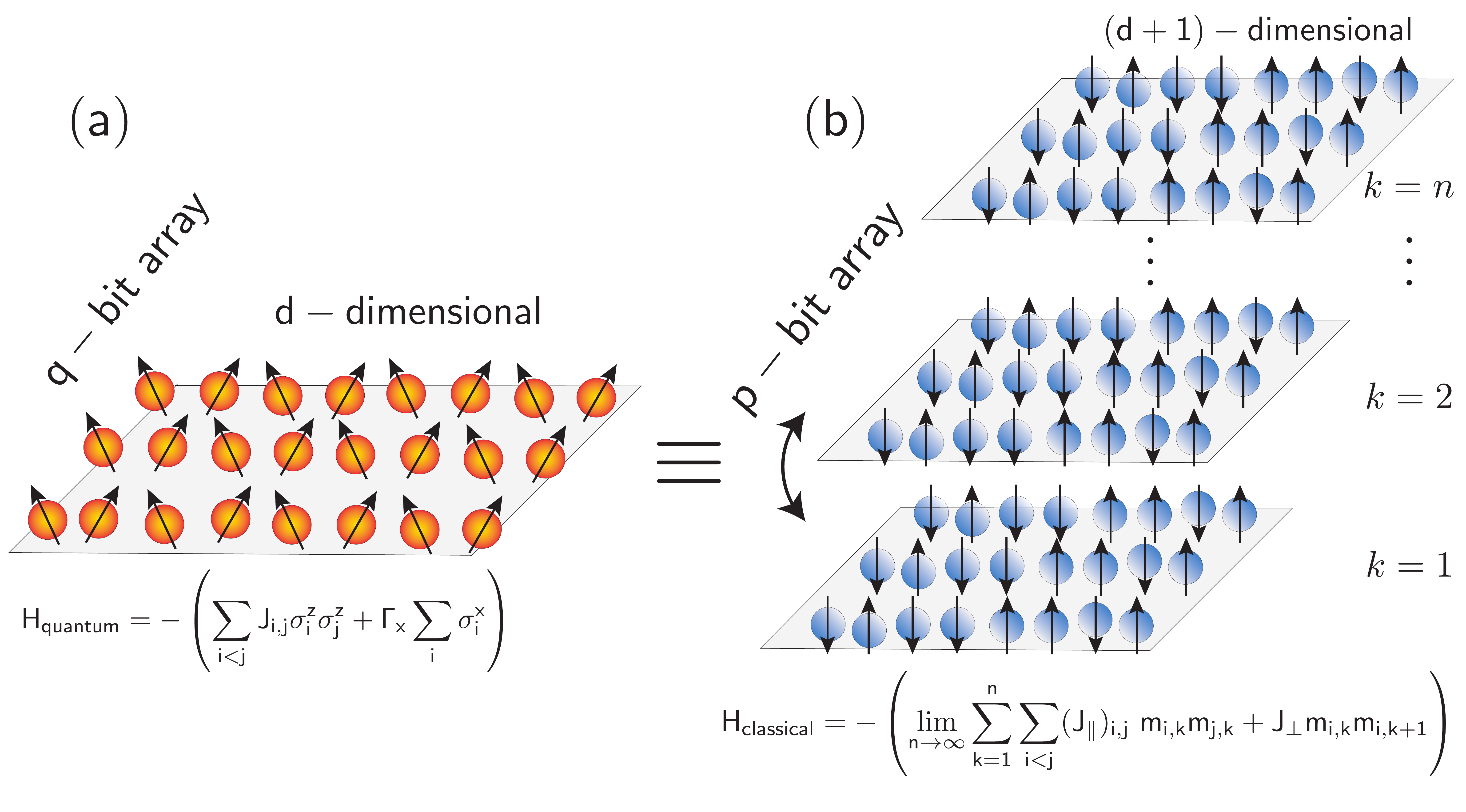}
\caption{\textbf{q-bit to p-bit mapping:} (a) A $d$-dimensional q-bit array described by the Transverse Ising Hamiltonian can be mapped to a $d$+1-dimensional p-bit array with $n$-replicas that are coupled in the vertical direction by the Suzuki-Trotter decomposition (The case for $d=2$ is illustrated). In this scheme, the replicas are always connected with periodic boundary conditions such that $m_{i,n+1}=m_{i,1}$. The many-body quantum and the corresponding classical Hamiltonian are shown where the operators $\sigma^z, \sigma^x$ of the quantum system are replaced with binary p-bits in the classical system with $m_{i,j}\in\{-1,+1\}$. Corresponding coupling terms are $(J_\parallel)_{i,j}=J_{i,j}/n$ and $ J_\perp = -1/(2\beta)\mathrm{log} \ \mathrm{tanh}(\beta \Gamma_x /  n)$.} 
\label{fi:fig1}
\end{figure*}
The basic building block of conventional digital electronics is the CMOS (Complementary Metal Oxide Semiconductor) transistor that is used to represent deterministic bits, that are either  0 or 1. Quantum computing, on the other hand, is based on q-bits that are coherent, delicate superpositions of  0 and 1. It is possible to define an entity intermediate between bits and q-bits  that are classical but probabilistic, which we call ``p-bits'' \cite{camsari2017stochastic}. It has been argued that just as three-terminal transistors provide a building block for large functional circuits, a three terminal realization of the p-bit can provide a building block for p-circuits \cite{behin2016building} reminiscent of the probabilistic computer described by Feynman in the same paper that helped launch the field of quantum computing \cite{feynman1982simulating}.

Such p-circuits can perform useful functions broadly relevant in the context of quantum computing  and machine learning \cite{camsari2018p}. For example, p-circuits can be used to perform classical annealing in hardware \cite{sutton2017intrinsic}, perform integer factorization by operating multipliers in an invertible mode \cite{camsari2017stochastic,pervaiz2017hardware}, just like quantum annealers that have been used for similar applications \cite{martovnak2004quantum,peng2008quantum}. In the machine learning context, p-bits can function as hardware accelerators for binary stochastic neurons \cite{ackley_learning_1985} that can be used to become efficient inference engines \cite{faria2018implementing,zand2018low}, or they can be used in an efficient calculation of correlations to accelerate learning algorithms,  an application area also discussed in the context of quantum computing \cite{bian2010ising,adachi2015application,liu2018adiabatic,amin2018quantum}. 

\subsection{Scope}

In this paper, we introduce an application of p-circuits to accelerate Quantum Monte Carlo (QMC) simulations of quantum systems based on the well-known Suzuki-Trotter decomposition that maps a $d$-dimensional quantum many body Hamiltonian to a $d$+1-dimensional classical Hamiltonian. This allows a quantum system to be emulated by a number of classical replicas that are interacting with each other \cite{suzuki1976relationship} (FIG.~\ref{fi:fig1}) and this approach is commonly used in software or high-level hardware simulations \cite{santoro2002theory,heim2015quantum,denchev2016computational,baldassi2018efficiency,hitachi_sqa_2017}. By contrast, we show that a compact, embedded MTJ-based coprocessor can speed up the simulation by several orders of magnitude. 

For a class of quantum Hamiltonians generally referred to as stoquastic Hamiltonians \cite{lidar_adiabatic_2018} that avoid the sign problem \cite{troyer2005computational} and are therefore amenable to efficient QMC simulation, it should be possible to build hardware accelerators using replicated p-bits to emulate the thermodynamics of q-bit networks. The number of p-bits required to emulate a given q-bit network is typically a factor of 25-100 larger, but this is offset by the relative ease of implementation. Three-terminal p-bits can be implemented at room temperature with Magnetoresistive Random Access Memory (MRAM) technology which is currently in production with hundreds of millions memory cells. Non-magnetic and completely digital implementations of p-bits are also possible  \cite{pervaiz2017hardware,pervaiz_weighted_2017} though they would require much larger energy and area \cite{zand2019composable} and while they can provide speed up over CPU/GPU implementations they would not achieve the potential speed up that can be obtained with the MTJ-based implementation.  

A highly efficient classical coprocessor made out of conventional p-bits could overcome fundamental difficulties associated with the low temperature operation of quantum annealers \cite{albash2017temperature} while operating almost as fast as physical annealers. For example, it has recently been shown that an optimized CPU-based simulated quantum annealing (SQA) implementation was $10^8$ times slower than a physical quantum annealer, even though it has shown similar algorithmic scaling on a model problem \cite{denchev2016computational}.

With appropriate magnet designs \cite{hassan2019low} individual p-bits can flip in a nanosecond or less so that with a million of them operating in parallel, we should have $\sim$ petaflips per second which is several orders of magnitude faster than existing digital implementations including parallelized GPU  \cite{fang2014parallel} and multi-core CPU implementations \cite{boixo_evidence_2014} that operate typically with $\sim$ 1-30 gigaflips per second.

It has also  recently been suggested \cite{albash2018demo} that among quantum annealing options, SQA  exhibits the best scaling properties, performing even better than experimental quantum annealers in some cases. As such, accelerating the software-based SQA with specialized hardware is a desirable goal and has led to recent interest in this area \cite{hitachi_sqa_2017,waidyasooriya2019opencl}. 

Another advantage of p-bit networks is that unlike q-bit networks they can be interconnected using conventional electronic devices such as GPUs or FPGAs. This could allow all-to-all connectivity beyond nearest neighbor coupling without requiring any special encoding \cite{lechner2015quantum,de2016simple}. Moreover, it should allow the implementation of arbitrary $k$-body interactions that are usually avoided by introducing ancillary bits to map them into 2-body interactions \cite{biamonte2008nonperturbative,jiang2018quantum}. 

\subsection{Organization of the paper}

\begin{figure*}[t!]
\includegraphics[width=0.85\linewidth]{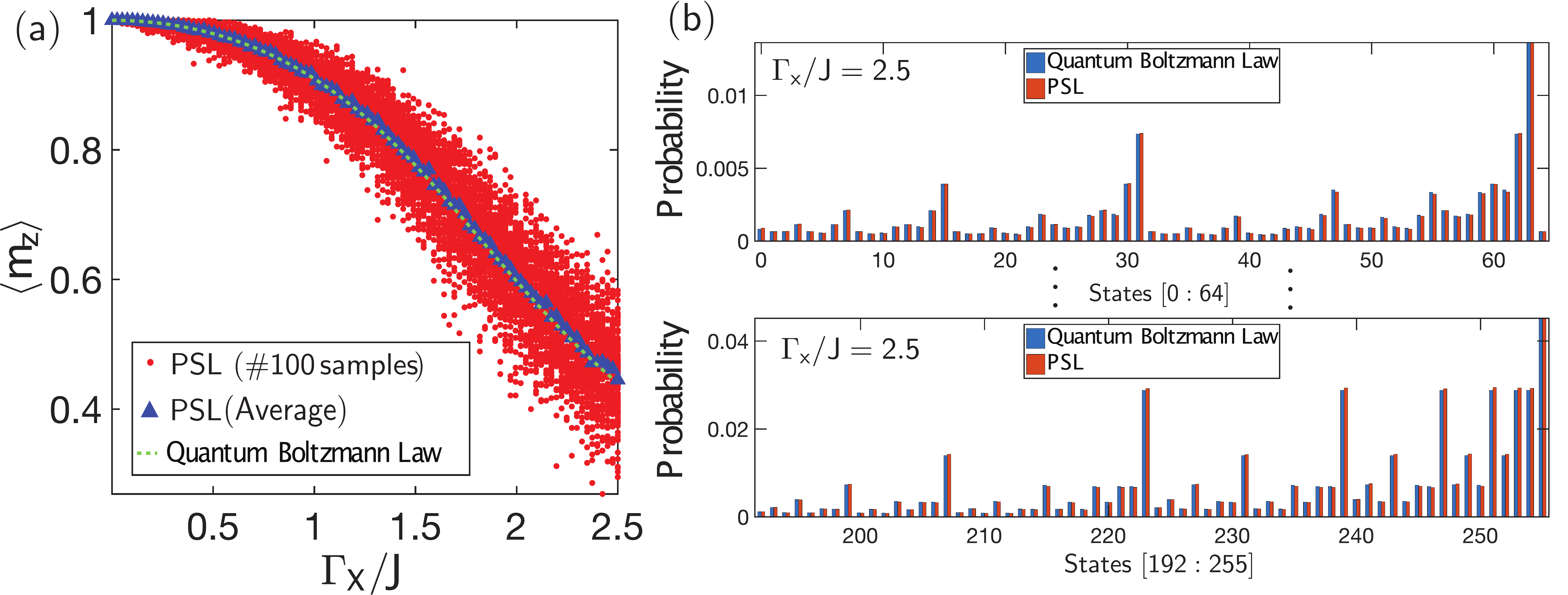}
\caption{\textbf{Exact quantum solution of a 1D Transverse Ising Hamiltonian vs Probabilistic Spin Logic (PSL):} (a) A 1D ferromagnetic linear chain ($J_{i,j}=+2$) with $M=8$ spins (nearest neighbor with periodic boundary conditions) described by the quantum Transverse Ising Hamiltonian (Eq.~\ref{eq:isingH}) is solved exactly, as a function of the transverse magnetic field ($\Gamma_x$) at an inverse temperature of  $\beta=10$. A symmetry breaking magnetic field in the $+\hat z$ direction is used, $\Gamma_z =1$, so that at $\Gamma_x=0$, all spins are pointing in the +$\hat z$ direction. The green dashed line is obtained by evaluating Eq.~\ref{eq:qboltzmann} as a function of $\Gamma_x$. The red dots represent 100 different PSL runs obtained with different RNGs, each running for $t_f=2000$ time steps. The blue triangles represent the average of PSL simulations and closely match the exact solution, establishing the accuracy of the quantum to classical mapping, with $n=250$ replicas. (b) A probability histogram of correlations of the form $\ket{\downarrow \downarrow \ldots \downarrow} =0, \ket{\uparrow \uparrow \ldots \uparrow}=255$  are obtained from PSL and Quantum Boltzmann Law at $\Gamma_x/J=2.5$ that corresponds to the last point of the $x$-axis in (a). Only a portion of the states are shown for clarity, states in between show essentially identical agreement.}  
\label{fi:fig2}
\end{figure*}

We start in Section \ref{sec:q2p}, with a description of the mapping from the q-bit network to the p-bit network, along with the behavioral equations describing the dynamics of the latter. These behavioral equations for p-circuits are similar to those used for stochastic neural networks and are often implemented in software for machine learning applications. However, a hardware implementation can provide a significant speed-up especially because it can allow parallel asynchronous operation under the right conditions.

Next in Section \ref{sec:model} we consider a common example of a stoquastic Hamiltonian, namely the Transverse Ising Hamiltonian \cite{PhysRevE.58.5355,pfeuty1970one}, commonly employed by quantum annealers \cite{johnson2011quantum}. We compare the exact quantum results for the averages and correlations with the results obtained from the p-bit network demonstrating the impressive accuracy  that can be achieved with a limited number of replicas. In Section \ref{sec:heis} we show another example, namely the ferromagnetic Heisenberg Model \cite{suzuki1976relationship,barma1978classical}, once again comparing the exact quantum results with probabilistic simulations of the p-bit network. 
In Section~\ref{sec:fac}, we show how Classical and Quantum Annealing can be performed using a network of p-bits. Finally in Section \ref{sec:p2mram} we present SPICE simulations of actual hardware implementations that can be built with existing Embedded Magnetoresistive RAM (eMRAM) technology  that has been under development by a number of foundries \cite{lin200945nm,song2016highly,globalMRAM}. Unlike standard eMRAM  where a non-volatile MTJ is carefully engineered  with a large energy barrier ($E_B \approx 40$-$60 \ k_BT$) so that the magnetization state is retained for a long time \cite{bhatti2017spintronics}, the free layer of the MTJ for the p-bit is designed as a thermally unstable magnet ($E_B \approx 0 \ k_BT)$ whose magnetization rapidly fluctuates in time in the presence of thermal noise \cite{camsari2017implementing}. Using full device-level SPICE simulations corresponding to the p-bit and a resistive interconnection matrix, we demonstrate that the correct quantum correlations can be obtained using this classical p-circuit which can be built with existing technology at room temperature. 

\section{q-bit to p-bit}
\label{sec:q2p}

Since the seminal work of Suzuki \cite{suzuki1976relationship}, it is well-known that a $d$-dimensional quantum many-body Hamiltonian can be mapped to a $d$+1-dimensional classical Hamiltonian applying the so-called Suzuki-Trotter decomposition \cite{trotter1959product,suzuki1976relationship}, which is used as a basis for  PIMC methods to simulate quantum annealing using classical computers \cite{santoro2002theory}. This decomposition results in the quantum system being mapped to a classical system with $n$ replicas that are coupled to each other. In this paper we consider two examples as described in the next two Sections, but the principles apply to stoquastic Hamiltonians in general.
 
Consider for example the Transverse Ising Hamiltonian in 1D written as \cite{pfeuty1970one}:  
\begin{equation}
 \mathcal{H}_{Q} \hspace{-3pt}=\hspace{-3pt}-\hspace{-3pt}\left(\sum_{i}^{M} J_{i,i+1} \sigma^z_i \sigma^z_{i+1} + \Gamma_x \sum_i^{M} \sigma^x_i + \Gamma_z \sum_i^{M}\sigma^z_i \right)
 \label{eq:isingH}
 \end{equation}
 
\noindent The Suzuki-Trotter mapping produces the following classical 2D Hamiltonian \cite{santoro2002theory}:
\begin{eqnarray}
\mathcal{H}_{C} &=& -\bigg(\lim_{n \to \infty }\sum_{k=1}^{n}\hspace{-1pt}\sum^{M}_{i=1} (J_{\parallel})_{i,i+1} \ m_{i,k} m_{i+1,k} + \gamma_z m_{i,k} \nonumber \\  
&& +J_{\perp} \  m_{i,k}m_{i, k+1}\bigg)
\label{eq:mappedH}
\end{eqnarray} 
where $(J_{\parallel})_{i,j}=J_{i,j}/n$, $n$ being the number of replicas, $\gamma_z  = \Gamma_z / n $ and the vertical coupling term is $J_{\perp}=-1/(2\beta)\mathrm{log \ tanh}(\beta \Gamma_x / n)$ and $m_{i,j} \in \{-1,+1\}$. Note how the quantum mechanical operators in Eq.~\ref{eq:isingH} have become classical spins in Eq.~\ref{eq:mappedH}. The mapping of Eq.~\ref{eq:mappedH} becomes exact in the limit of infinite replicas ($n\rightarrow \infty$) however, for finite replicas the error scales as $O(1/n^2)$ \cite{heim2015quantum}  and can be made  arbitarily small by choosing an appropriate number of replicas.

\subsection{Behavioral model for p-bits}
\label{sec:pbit}

\begin{figure*}[t!]
\includegraphics[width=0.85\linewidth]{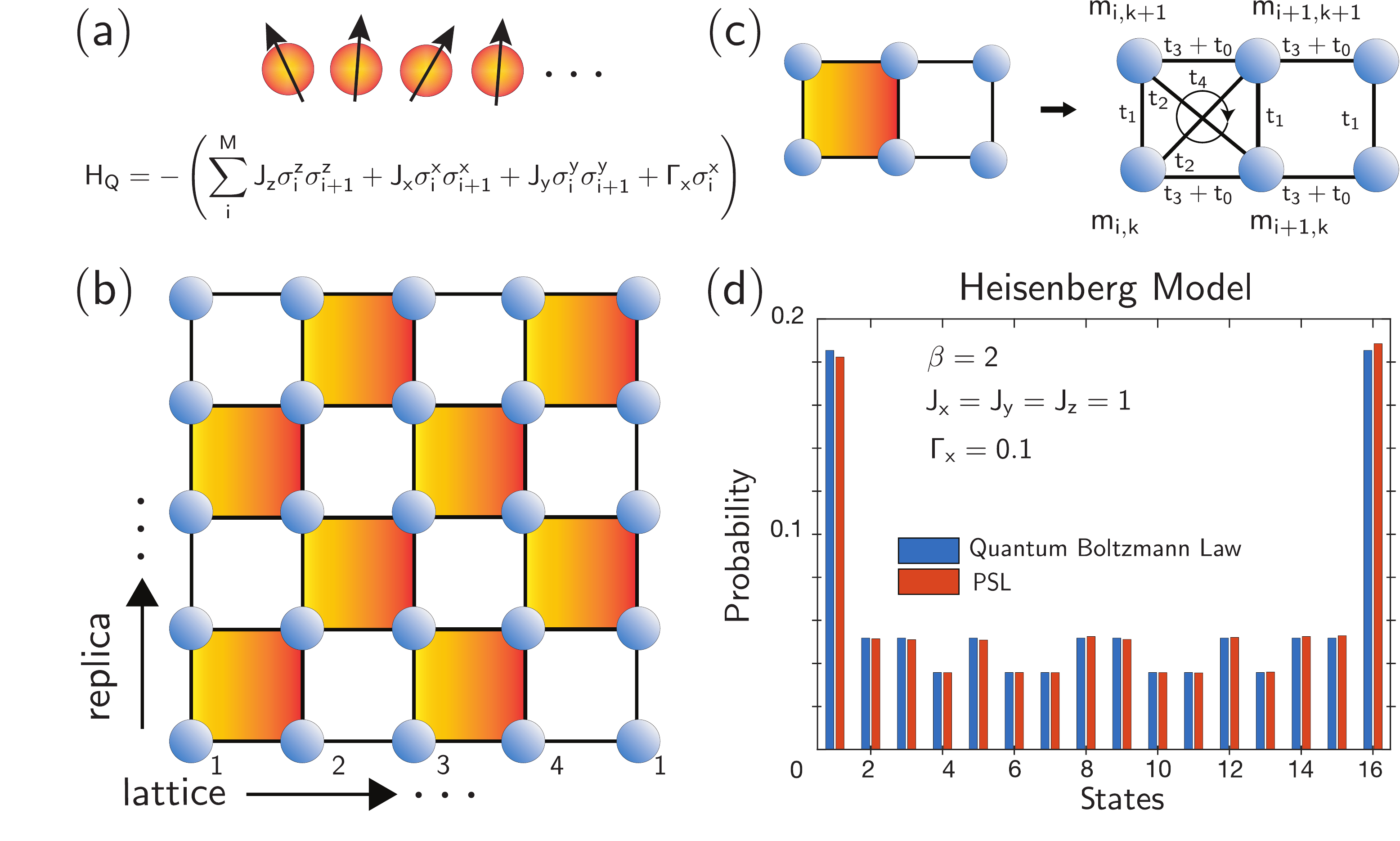}
\caption{\textbf{1D Heisenberg Model:} (a) A 1D Heisenberg model ($M=4$ spins) with a transverse magnetic field  is considered. (b) The chessboard lattice corresponding the 2D classical mapping of the model  with periodic boundary conditions in the replica and lattice directions. (c) The interaction terms within the shaded and unshaded unit cells are shown. Within shaded cells, all neighbors are coupled vertically, diagonally and horizontally ($t_1, t_2, t_3$) in addition to a 4-body interaction energy term $(t_4)$ that involves the product of all four spins. Terms that arise from the diagonal part of the Hamiltonian, $H_0$, result in additional horizontal interactions ($t_0$) in both the shaded and unshaded cells. (d) Probability histogram corresponding to a $M=4$ spin ferromagnetic Heisenberg Model. The histogram is obtained by solving the Quantum Boltzmann Law (Eq.~\ref{eq:qboltzmann}) with the parameters shown in the inset and by solving behavioral PSL equations but with a modified Eq.~\ref{eq:synapse} to account for a 3-body term arising from the 4-body interaction. For the PSL simulation 20 replicas are used with $t_f =10^7$ time steps. Samples taken from different replicas are considered independent and reduced to 16 probabilities to be compared with the original quantum system.} 
\label{fi:fig3}
\end{figure*}

The classical system expressed by Eq.~\ref{eq:mappedH} can be represented by p-circuits that are built out of p-bits. There are two central equations that are used to describe p-bit networks \cite{camsari2017stochastic}: 
\begin{subequations}
\begin{equation}
m_i (t+1) = \mathrm{sgn} \big[ {r + \mathrm{tanh} \ \beta I_i (t) \big]}
\label{eq:pbit}
\end{equation}
where $t$ is dimensionless time that is incremented one at a time, $r$ is a random number uniformly distributed between $-$1 and +1 and $r$ at each time step is uncorrelated with the $r$ chosen at the previous step. $\beta I_i$ is the dimensionless current to each p-bit, where $\beta$ is the inverse temperature. $I_i$ in general, is calculated according to,
\begin{equation}
{I_i}(t) \equiv -\frac{\partial \mathcal{H}_C}{\partial m_i}  \nonumber
\end{equation}
which in the present case, becomes:
\begin{equation}
I_i(t)=\bigg( b_i+\sum_j{W_{ij}m_j (t) } \bigg) 
\label{eq:synapse}
\end{equation}
\label{eq:PSL}
\end{subequations}

\noindent where $W_{ij}$  is the interconnection matrix and $b_i$ is the bias term. We refer to Eq.~\ref{eq:PSL} as Probabilistic Spin Logic (PSL) equations and note that these equations are essentially the same as those discussed in the context of stochastic neural networks such as Boltzmann Machines, developed by Hinton and colleagues \cite{ackley_learning_1985}. 

It is important to note that  while Eq.~\ref{eq:synapse} is a linear synapse that typically arises from quadratic Hamiltonians with 2-body interactions,
specially designed digital CMOS circuits can be used to implement more complicated interactions arising from cost functions such as generalized Hopfield models with $k$-body interactions \cite{gardner1987multiconnected,seki2015quantum}.  Such a flexibility of implementing complicated interactions could be a key advantage for hardware p-circuits. 

\subsection{PSL dynamics}
PSL equations can be updated to approximate the steady state joint probability density for any $W$ matrix, symmetric or asymmetric. For symmetric $W$ matrices, the joint probability density is simply expressed by the classical Boltzmann Law, $\rho(\{{m}\}) \propto{\exp[-\beta E(\{m\})]}$,  where $E$ is the energy for a given configuration $\{m\}$, $E = 1/2 \  {m^T} [W] {m}$. There are two important conditions regarding the updating of Eq.~\ref{eq:PSL}. First, Eq.~\ref{eq:synapse} needs to be calculated much faster than Eq.~\ref{eq:pbit} for proper convergence \cite{pervaiz2017hardware}, a requirement particularly relevant for hardware implementations. Second, Eq.~\ref{eq:pbit} needs to be updated sequentially, as in Gibbs
sampling \cite{geman1984stochastic}. The requirement of sequential updating prohibits parallelization in software implementations, except in special cases such as restricted Boltzmann machines where the lack of intralayer connections between ``visible'' and ``hidden'' layers allows each layer to be updated in parallel \cite{hinton2012practical}. For asynchronous  hardware implementations, however, a clockless operation seems to satisfy the requirement of sequential updating naturally \cite{pervaiz2017hardware,sutton2017intrinsic}.

\section{Transverse Ising Hamiltonian}
\label{sec:model}
For the 1D Transverse Ising Hamiltonian (Eq.~\ref{eq:isingH}), we assume periodic boundary conditions such that  $\sigma^z_{M+1}=\sigma^z_1$. $\Gamma_x$ is the local transverse magnetic field and $\Gamma_z$ is a local $z$-directed magnetic field.  Eq.~\ref{eq:isingH} can be constructed by first writing each term,  $\sigma_i$, as a $2^M \times 2^M$ matrix followed by ordinary matrix multiplication for each product term. These terms are written in terms of $2\times 2$ Pauli spin matrices ($\varsigma^{x,y,z}$) at the $j^{\rm th}$ lattice point as $\sigma_{j}= I \otimes I \otimes \ldots  \otimes  {\mathbf{\varsigma}}\otimes \ldots \otimes I \otimes I$ where $I$ is the 2$\times 2$ identity matrix and $\varsigma$  is the Pauli spin matrix  at  the $j^{\rm th}$ term in the product.

 \subsection{Quantum Boltzmann Law}
In principle, Eq.~\ref{eq:isingH} can be exactly solved for any quantity of interest as a function of temperature and all other parameters  $J$ and $\Gamma$,  from the principles of quantum statistical mechanics \cite{kadanoff1962quantum}: 
\begin{equation}
\langle S \rangle  =  \frac{\mathrm{Tr}. \  [ S_{op}  \exp (- \beta \mathcal{H}_{Q})]}{\mathrm{Tr}. \  [\exp(-\beta \mathcal{H}_Q)]}
\label{eq:qboltzmann}
\end{equation} 
where $\beta \equiv 1/k_BT$ is the ``inverse temperature" (as defined in Eq.~\ref{eq:pbit}) and we have chosen to use a unit system in which $k_B = 1$. $S$ is the quantity we wish to calculate with a corresponding operator $S_{op}$. In practice, directly solving Eq.~\ref{eq:qboltzmann} becomes intractable due to the exponential dependence of the Hamiltonian ($2^M \times 2^M$) to the size of the problem ($M$). Due to its similarity to the classical Boltzmann Law \cite{feynman2011feynman}, we refer to Eq.~\ref{eq:qboltzmann} as the ``Quantum Boltzmann Law'' throughout this paper and solve it for small 1D systems. To obtain numerically stable results at low temperatures (high $\beta$), we first diagonalize the Hamiltonian and subtract the ground state energy from the diagonals, without changing any observable quantities.

\subsection{Averages and correlations} 
In FIG.~\ref{fi:fig2}a we calculate the average $z$-magnetization of a 1D ferromagnetic ($J_{ij}=+2$) chain with $M=8$ spins, as a function of a transverse magnetic field. The average $z$-magnetization, $\langle m_z \rangle$,  is obtained by the operator  $\sigma^z = \sum {\sigma^z_j}/M$  where $\sigma^z_j$ provides the net $z$-spin, $\ket\uparrow - \ket\downarrow$, at  site $j$. To break the symmetry of $m_z=\pm 1$ at low temperatures $(\beta=10)$ we introduce a $+\hat z$-directed magnetic field. As the transverse magnetic field increases, $\langle m_z \rangle$ gradually decreases, while $\langle m_x\rangle$ (not shown) increases, as spins become aligned with the transverse magnetic field. Incidentally, the reverse process, starting from a large $\Gamma_x$ at a low temperature and slowly decreasing it to find the ground state of a complicated spin-glass, is commonly used in quantum annealing algorithms \cite{heim2015quantum}.

FIG.~\ref{fi:fig2}b shows the probabilities of correlated states at a given temperature and transverse field expressed as decimal numbers. This is done by first converting the states to binary numbers such that $\uparrow$ denotes +1 and $\downarrow$ denotes 0 and then converting the full state into a decimal number, for example the all down state $\ket{\downarrow \downarrow \ldots \downarrow}$ corresponds to 0, and the all up state $\ket{\uparrow \uparrow \ldots \uparrow}$ corresponds to 255 and so on. There are $2^8=256$ such states, each with a given probability obtained from Eq.~\ref{eq:qboltzmann}. These correlated states are calculated by first constructing an operator for the probability of finding a $\ket\uparrow$ state at a given site, $P_j (\ket{\uparrow}) = (I + \sigma^z_j)/2$ where $I$ is the $2^M \times 2^M$ identity matrix. Similarly, $P_j (\ket{\downarrow}) = (I - \sigma^z_j)/2$. Using these operators, any correlation of the form $\ket{\downarrow \uparrow \ldots \uparrow}$ can be calculated from the corresponding composite operator:

\begin{equation}
P(\downarrow \uparrow \ldots \uparrow) = P(\downarrow)P(\uparrow) \ldots P(\uparrow) = \prod_{k=1}^{M} P_k
\end{equation}
There are 256 such operators and Eq.~\ref{eq:qboltzmann} can be used for each of them to obtain a probability for each state for any $J, \Gamma, \beta$. FIG.~\ref{fi:fig2}b shows these probabilities at a chosen parameter combination and they are in agreement with  results obtained from a simulation of p-bits, as we next explain in Section~\ref{sec:q2p}. Note that this joint probability density contains all statistical information in the system, as averages and other correlations of interest can be calculated from it, for example one can obtain $\langle m_z\rangle$ by weighting each state by the net $z$-spin they contribute to the average.

\subsection{PSL vs Quantum Boltzmann Law}

With this picture, the mapped classical Hamiltonian with $n$ replicas described in  Eq.~\ref{eq:mappedH} is used to obtain a consolidated $[W]$ matrix that is of size $ (M n)\times (M n)$ to be used in Eq.~\ref{eq:PSL}. FIG.~\ref{fi:fig2} shows the equivalence of the PSL implementation of the Transverse Ising Hamiltonian to the exact quantum many-body description for a 1D-chain with $M=8$ spins. Note that the p-bit mapping can be applied to much larger spin systems, but an exact solution by  Eq.~\ref{eq:qboltzmann} quickly becomes intractable. We investigate the average $z$-spin of this ferromagnetic chain at a constant temperature ($\beta=10$)  as a function of the transverse magnetic field, $\Gamma_x$. A symmetry breaking field (to favor a +1 order) of $\Gamma_z=1$ is applied. As expected, the exact result shows how the average $z$-spin becomes disordered. The PSL results for a $n=250$ replica system reproduce this behavior. The $z$-spin average is obtained by taking an average over the length of the chain, as well as over each replica. The final average (for a given red dot) is recorded at the end of $t_f=2000$ dimensionless time steps. Since a single stochastic point is recorded at the end of $t_f$, for each $\Gamma_x$ point, we observe a variance in the final results, however averaging over 100 different simulations for the same system, we get a very close match to the exact solution.

In FIG.~\ref{fi:fig2}b, the full joint probability density for the classical system is obtained from a PSL simulation that is run for $t_f=10^{5}$ dimensionless time steps. The state of each replica with 8-spins is converted into a binary number at each time step, as in the exact solution, and then collected over all replicas. The striking agreement with PSL and the Quantum Boltzmann Law in FIG.~\ref{fi:fig2} establishes the faithful mapping of the quantum system to the classical system, from the behavioral PSL equations Eq.~\ref{eq:PSL}.

\section{Ferromagnetic Heisenberg Model}
\label{sec:heis}

Before proceeding to a hardware implementation showing how replicated networks of p-bits can be built by existing nanodevices, we show another example of a stoquastic Hamiltonian that can be represented by p-bits. The Heisenberg Hamiltonian in 1D in the presence of a transverse magnetic field can be written as: 
\begin{equation}
\mathcal{H}_Q \hspace{-3pt} = -\hspace{-1pt} \hspace{-1pt}\left(\hspace{-1pt}\hspace{-2pt} \sum_{i}^{M} J_z \sigma^z_i \sigma^z_{i+1} \hspace{-3pt}+ \hspace{-3pt} J_x \sigma^x_i \sigma^x_{i+1}  \hspace{-3pt}+ \hspace{-3pt}J_y \sigma^y_i \sigma^y_{i+1} \hspace{-3.5pt}+ \hspace{-3.5pt} \Gamma_x \sigma^x_i \hspace{-4pt}\right) \hspace{-3pt}\end{equation}

Following \cite{suzuki1976relationship,bravyi2014monte,barma1978classical}, we apply the Suzuki-Trotter transformation to this system and obtain the chessboard lattice that is shown in Fig.~\ref{fi:fig3}b with shaded and unshaded unit cells. The interactions terms for this hardware neural network are shown in Fig.~\ref{fi:fig3}c. For the shaded unit cells all two-body interactions ($t_1, t_2, t_3,$) exist in addition to a 4-body interaction ($t_4$) that involves the product of each spin. The two-body interactions can be implemented using a linear synapse of the form of Eq.~\ref{eq:synapse} but the 4-body interaction requires a non-linear synapse that computes the input terms that are products of three neighboring spins. The interaction terms $t_0$ arise due to the diagonal parts of the quantum system, as in the case of the Transverse Ising Hamiltonian, and exists for both the shaded and unshaded unit cells shown in Fig.~\ref{fi:fig3}c. The detailed derivations of all interaction terms are shown in Appendix A. 

In Fig.~\ref{fi:fig3}d, we show a simulation of the  classical system using behavioral PSL equations and compare this to the exact solution as before. We choose a set of parameters,
$J_x = J_y = J_z = 1$ that corresponds to the ferromagnetic Heisenberg Model with a small transverse magnetic field in the $x$-direction, such that all off-diagonal terms in the $\exp(-\beta \mathcal{H_Q})$ are \textit{positive}, hence making this system \textit{stoquastic}  \cite{bravyi2014monte}. In this small example with $M=4$ spins, we observe good agreement between the mapped system and the exact solution.  

\section{Factorization as Inverse Multiplication} 
\label{sec:fac}

\begin{figure*}[ht!]
\includegraphics[width=0.75\linewidth]{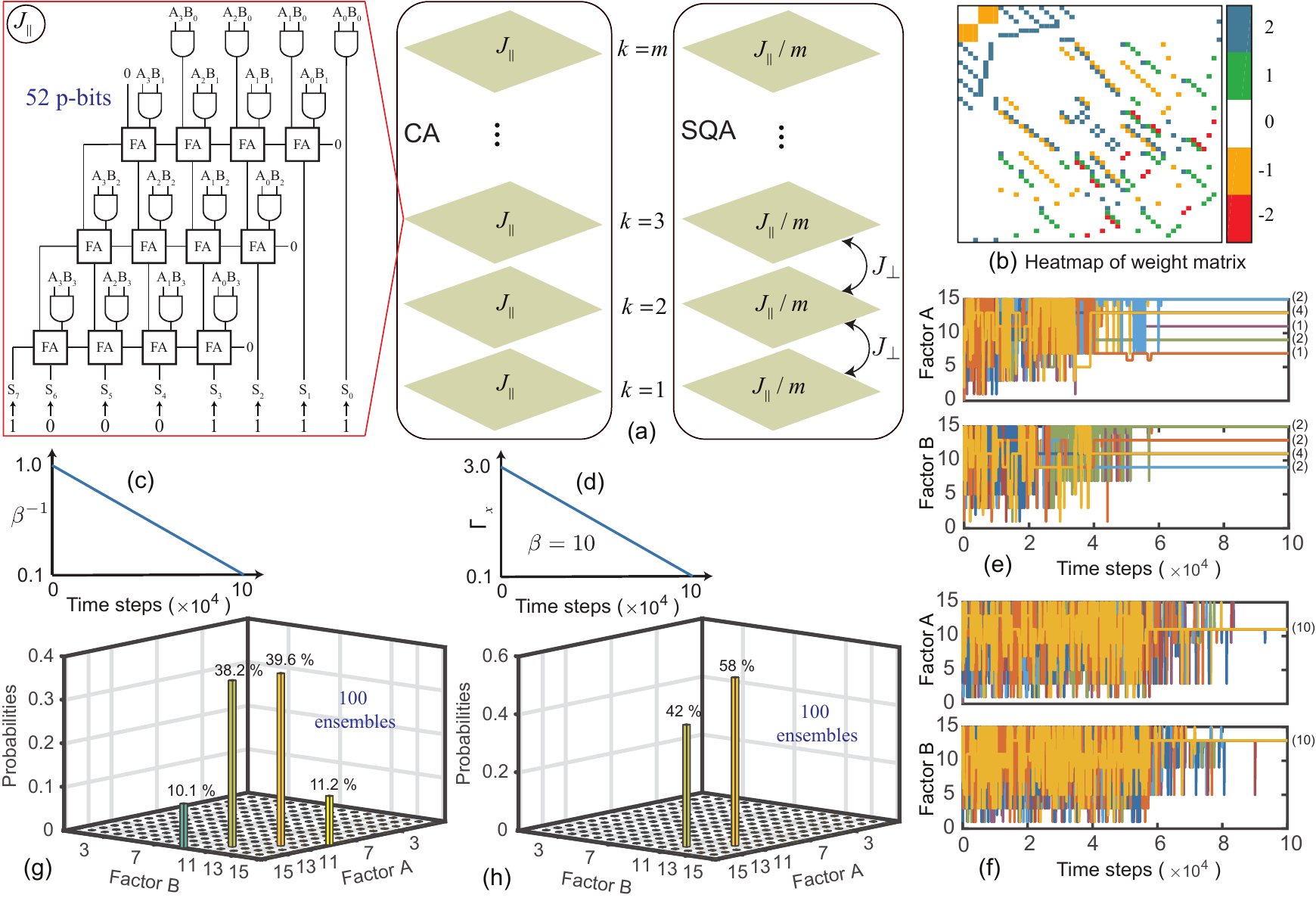}
\caption{\textbf{Classical versus quantum annealing with p-circuits:}  (a) An 8-bit binary multiplier designed as a p-circuit. When operated in invertible mode this circuit functions as a factorizer. In the classical annealing (CA) scheme, this circuit is replicated $m$ times ($m=10$) without any interaction between replicas but to collect equal statistics as simulated quantum annealing (SQA). For SQA, the weight matrix is obtained according to the Transverse Ising mapping as discussed. (b) The heatmap of the weight matrix $J_{\parallel}$ shows that the connections among p-bits in the invertible multiplier circuit are  sparse and discrete. (c) Classical annealing schedule: $\beta^{-1}$  is linearly decreased from 1 to 0.1. (d) In SQA, the transverse field ($\Gamma_x$) is linearly reduced from 3 to 1 while $\beta=10$. (e) The time evolution of CA: The numbers inside parenthesis on the right shows the multiplicity of replicas at a particular value. (f) Time evolution for SQA: All replicas find the right factors at the end annealing. (g-h) Histograms are calculated by averaging the results over 100 ensembles.} 
\label{fi:fig4}
\end{figure*}

So far we have shown probabilistic emulation of quantum systems in equilibrium without performing annealing. In Fig.~\ref{fi:fig4}, we show how classical and quantum annealing can be performed by the probabilistic coprocessor using Eq.~\ref{eq:pbit} and Eq.~\ref{eq:synapse}. We choose the problem of integer factorization by expressing a p-circuit that performs binary multiplication using Full Adders and AND gates. This structure is similar to the factorization descriptions in \cite{camsari_stochastic_2017,pervaiz2017hardware,traversa2017polynomial,smithson2019efficient}. We improve our previous design \cite{camsari_stochastic_2017,pervaiz2017hardware} by eliminating nodes that are connected to each other, for example if a Full Adder carry out is connected to the carry in of another Full Adder, these two nodes are combined into a single node so that the p-bit in this node receives the sum of the inputs that each node receives. This allows us to reduce the problem size and exhibits better scaling for the inverse multiplier. 

Fig.~\ref{fi:fig4} compares Classical Annealing (CA) with simulated Quantum Annealing (SQA) with p-circuits. For classical annealing the temperature $\beta^{-1}$ is linearly decreased from 1 to 0.1, while for simulated quantum annealing the inverse temperature is fixed at $\beta=10$ but the transverse magnetic field is linearly reduced from 3 to 0.1. For these parameters, we observe that for the 8 bit multiplier, the SQA seems to perform better than CA as SQA finds the correct factors with 100\% probability with 58\% probability of finding (11,13) and 42\% probability of finding (13,11) while CA finds the correct factors with 77.8\% probability out of 100 samples.  While we note that SQA seems to work better for this particular set of parameters, we have not attempted to optimize the parameters for CA or SQA. In SQA $m$-replicas of the original system is needed to map the Ising Hamiltonian to the corresponding Transverse Ising Hamiltonian. To make a fair comparison between SQA and CA in terms of the statistical samples being used, we performed CA with the same number of replicas but they are not interacting with each other. With a sophisticated synapse design that would allow replica swapping, replicas in the CA mode can be held at different temperatures for parallel tempering algorithms \cite{zhu2015efficient} but we do not explore this further. Our purpose has been to show that an autonomously operating p-circuit can be used to perform CA and SQA with \emph{physical replicas} to speed up these algorithms as we show in the next section. 

\section{p-bit to Stochastic MRAM}

\begin{figure*}[t!]
\includegraphics[width=0.99\linewidth]{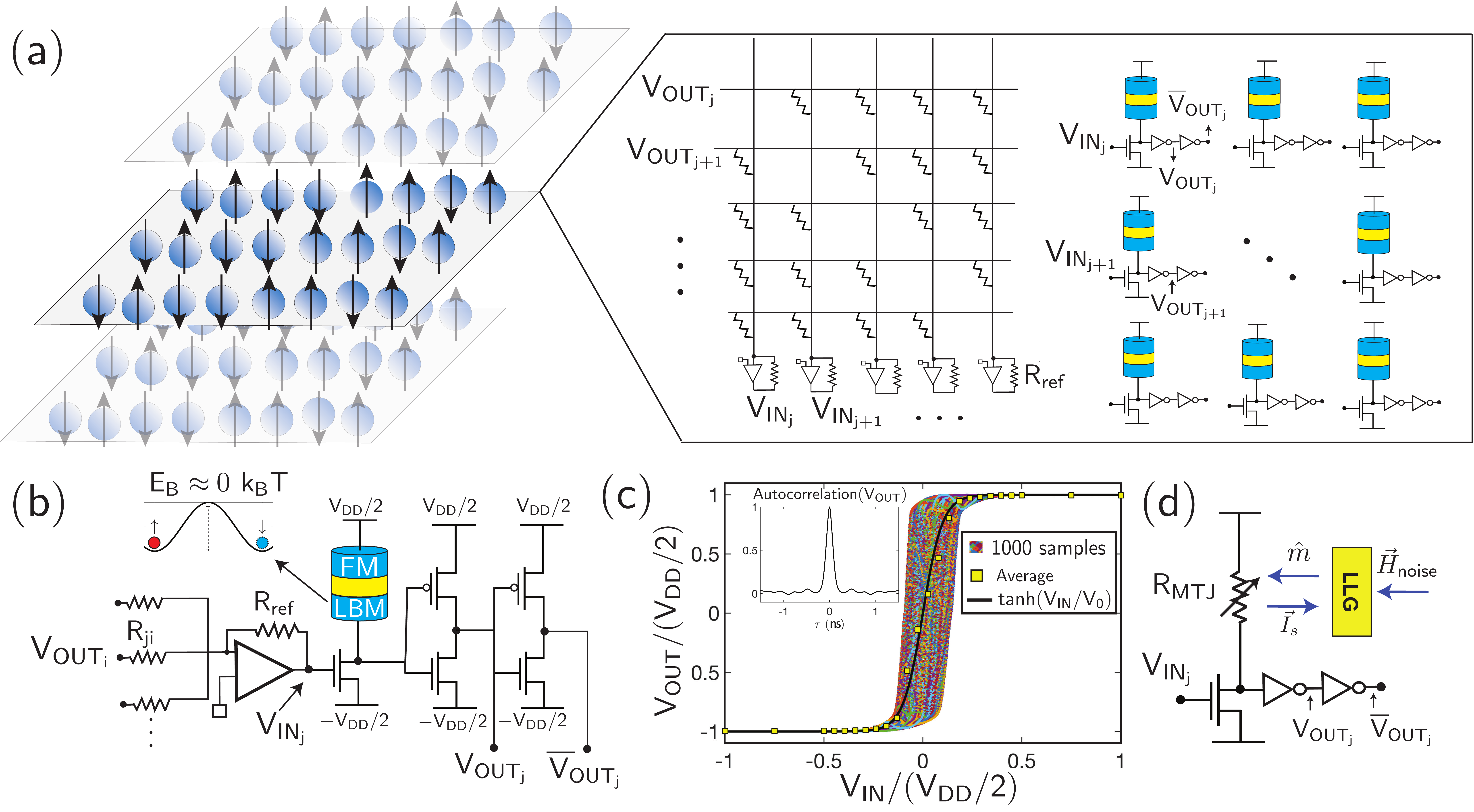}
\caption{\textbf{p-bit to Stochastic eMRAM:}  (a) Each replica in the classical system is represented by a hardware neural network involving p-bits (neurons), interconnected by a resistive network (synapse). The outputs of the p-bits are weighted by the resistive network to become inputs to each other. Bias terms are added as fixed external voltage sources. (b) Detailed circuit schematics of a given p-bit and synapse following Ref.~\cite{camsari2017implementing}: The outputs are collected by a fast operational amplifier (assumed ideal in circuit simulations). Fixed layer ferromagnet (FM) is a stable magnet with a large energy barrier, while the free layer is a circular low-barrier magnet (LBM) ($E_B \approx 0 \ k_BT$) whose magnetization fluctuates in the presence of thermal noise. (c) SPICE simulations for the input-output characteristics of the p-bit:  The results for 1000 p-bits where the input voltage is swept from $ -V_{\rm DD}/2$ to +$V_{\rm DD}/2$, in $t_{sim}=1$ ns (Inset shows the autocorrelation of the p-bit at $V_{\rm IN}=0$). Each p-bit has a randomized resistance due to the random magnetization of the free layer, showing a range of outputs bounded by the parallel and anti-parallel resistance of the MTJ. Ensemble averaged output for 1000 samples at a given input voltage ($t_{\rm sim}=2$ ns for each sample)  shows a tanh($V_{\mathrm{IN}}/V_0$) behavior. (d) The circuit model that self-consistently solves the stochastic LLG equation with the MTJ and transistor models. $\vec{I}_s$ is the spin-current exerted on the free layer due to the  current polarized by the fixed layer, $\hat{m}$ is the instantaneous magnetization and $\vec{H}_{\rm noise}$ is the thermal noise field.} 
\label{fi:fig5}
\end{figure*}

\begin{figure}[t!]
\centering
\includegraphics[width=0.80\linewidth]{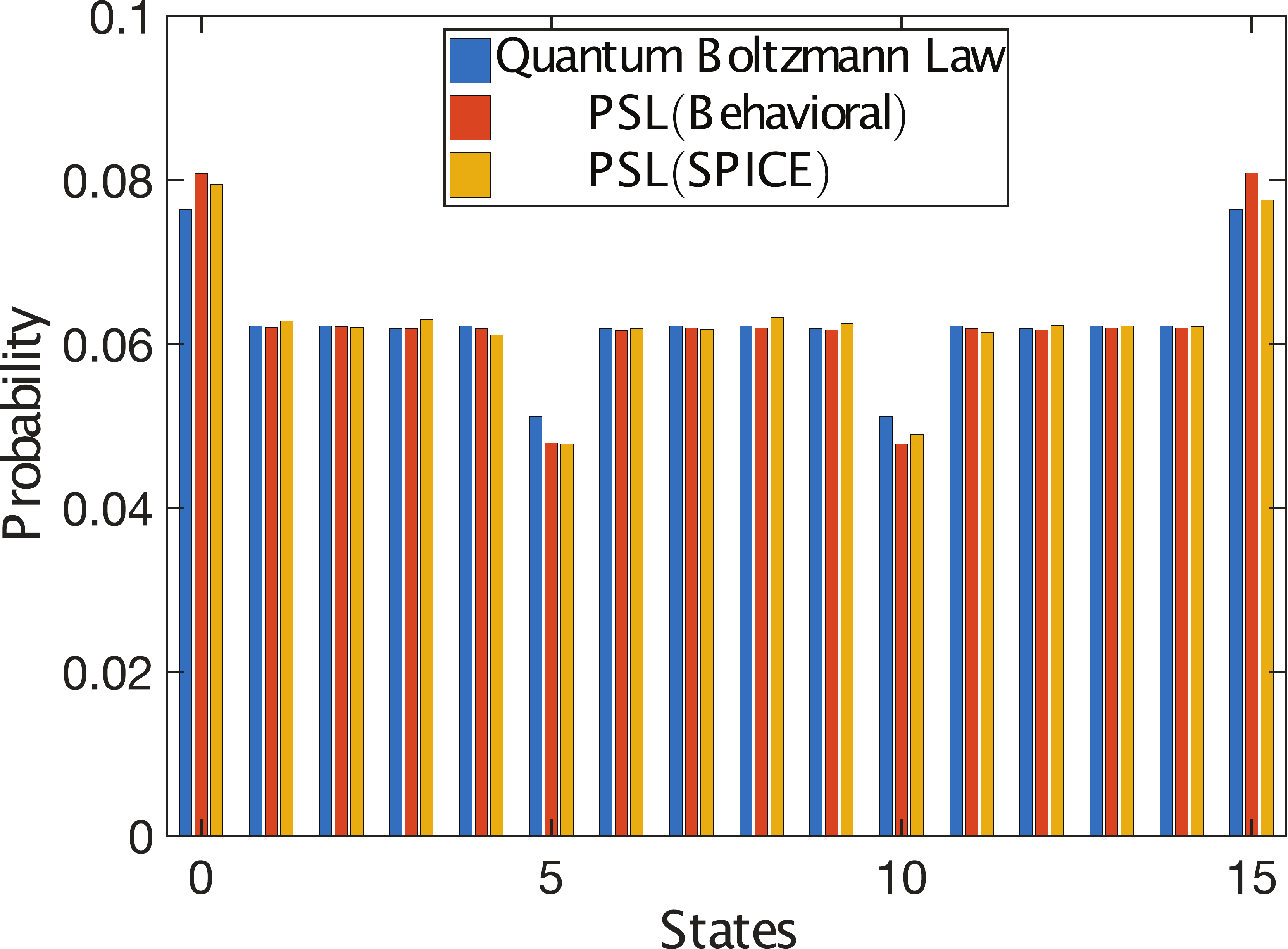}
\caption{\textbf{Full circuit simulation of a 4-spin chain with 10 replicas:} 1D classical Ising chain of $M=4$ spins is simulated in SPICE with full device models for the stochastic MRAM-based p-bit and a resistive interconnection matrix and compared with PSL and Quantum Boltzmann Equation. $\beta=0.5$, $\Gamma_x=10$ and $J_{i,j}=+1$. The joint probability density is expressed in decimal numbers similar to the previous examples.} 
\label{fi:fig6}
\end{figure}

\label{sec:p2mram}
We now show how the behavioral p-bit model can be represented by a stochastic neural network in hardware (FIG.~\ref{fi:fig5}a). Each replica in the classical system consists of p-bits that are interconnected to each other with a resistive network (synapse), a typical architecture often used in many hardware neural networks \cite{yu2011electronic,hu2016dot} though for more complicated systems involving $k$-body interactions ($k>2$), standard electronic devices such as FPGA's could also be used for this purpose, for example as in Ref.~\cite{mcmahon2016fully}. The extra dimension added by the Suzuki-Trotter transformation would increase the synaptic complexity but for sparse quantum networks, this transformation would only slightly increase the fan-in of each p-bit in the classical network. 

We assume that the weighted summation is carried out by ideal operational amplifiers. The replicas are also connected in the vertical direction (not shown in FIG.~\ref{fi:fig5}) with nearest neighbor coupling according to the coupling coefficient  $J_\perp$. 

In the case of quantum annealing, the vertical resistors need to be reconfigurable, therefore they need to be designed differently compared to the fixed resistors that represent the transverse coupling $(J_\parallel)_{i,j}$. In our device level examples, we use fixed resistors in order to  establish the equivalence between the classical and quantum systems and have not performed annealing. 

\subsection{Network parameters}
The device equations for the synapse and the p-bit shown in FIG.~\ref{fi:fig5} are given as \cite{camsari2017implementing,faria2018implementing}:  
\begin{eqnarray}
&&V_{\mathrm{OUT}_j} /(V_{\mathrm{DD}}/2) = \mathrm{sgn} [ r + \mathrm{tanh}(V_{\mathrm{IN}_j}/V_0)] \label{eq:mtj_pbit}  \\  [1em] 
&& V_{\mathrm{IN}_j} = \sum_i  \frac{{R_\mathrm{{ref}}}}{R_{ji}} \overline{V}_{\mathrm{OUT}_i} \label{eq:opamp}
\end{eqnarray}
Eq.~\ref{eq:mtj_pbit} and Eq.~\ref{eq:opamp} are combined with the PSL equations, Eq.~\ref{eq:PSL}, to obtain the following equations that map the behavioral PSL equations to physical  parameters:
\begin{equation}
m_j = \frac{{V}_{\mathrm{OUT}_j}}{(V_{\mathrm{DD}}/2)}, \quad W_{ji} = \frac{R_{0}}{R_{ji}}, \ \beta =\frac{V_{\mathrm{DD}} R_{\mathrm{ref}}}{2 V_0 R_0}
\end{equation}
where $R_0$ is a unit resistor that is used to electrically change the inverse temperature $\beta$, and $V_0$ is a transistor dependent parameter ($\approx 40$ mV) that defines the stochastic window of the p-bit (FIG.~\ref{fi:fig5}c). Depending on the sign of the interconnection, $W_{ij}$, the non-inverted output $V_{{\rm OUT}_j}$ or the inverted output $\overline{V}_{{\rm OUT}_j}$ is used for the synaptic connections.
 
\subsection{Device models}
The 1T/1MTJ p-bit is modeled by combining a 14 nm-High Performance FinFET model from the open source Predictive Technology Models (PTM) \cite{predictive_tech} with a stochastic Landau-Lifshitz-Gilbert (sLLG) solver implemented in SPICE \cite{torunbalci2018modular}, following the design described in \cite{camsari2017implementing} (FIG.~\ref{fi:fig5}d). The MTJ is modeled as a simple conductor whose conductance depends on the instantaneous magnetization $m_z(t)$, provided by the sLLG such that 
\begin{equation}
G_{\rm MTJ}(t)=G_0\left[1+m_z(t) \frac{R_{AP}-R_P}{R_{AP}+R_P}\right]
\end{equation}
     where $R_P$ and $R_{AP}$ are the parallel and antiparallel resistance of the MTJ and $G_0=(R_{AP}^{-1}+R_P^{-1})/2$. We use an experimentally measured value for the tunneling magnetoresistance (TMR) $=(R_{AP}-R_{P})/R_P=110\%$ after Ref.~\cite{lin200945nm}. $G_0$ is set equal to the transistor resistance at $V_{\mathrm{IN}}=0$ to produce a symmetric sigmoid with no offsets, in this case $G_0^{-1}= 23.4 \rm\  k\Omega$.  The free layer is assumed to be a circular low barrier nanomagnet \cite{cowburn_single-domain_1999,debashis2018designing}  with a diameter of 22 nm and thickness of 2 nm and a saturation magnetization of  $M_s=1100$ emu/cc, with a damping coefficient $\alpha=0.01$, typical parameters for CoFeB \cite{sankey2008measurement}.  

The time dependent magnetization is obtained by solving sLLG equation in the monodomain approximation \cite{sun2000spin}:
\begin{subequations}
\begin{align}
&(1+\alpha^2)\frac{d\hat m}{dt} = -|\gamma|\left(\hat m \times \vec{H}\right) - \alpha |\gamma| \left(\hat m \times \hat m \times \vec{H}\right)\nonumber \\ &+  \frac{1}{q  N}\left(\hat m \times \vec{I}_{S} \times \hat m\right)  + \frac{\alpha}{q N} \left(\hat m \times \vec{I}_{S}\right)
\label{sLLG}
\end{align}
\end{subequations}
  $\gamma$ is the electron gyromagnetic ratio, $q$ is electron charge and $N$ is the number of Bohr magnetons ($\mu_B$) in the volume of the magnet, $N= M_s\mathrm{Vol.} / \mu_B$.  $\vec{H}$ contains the external magnetic and internal anisotropy fields of the magnet as well as the noise field. In the case of a circular nanomagnet without an easy-axis anisotropy, the total internal magnetic anisotropy becomes $\vec{H_m}= - 4\pi M_s m_x \hat x$, where $z$-$y$ is the easy plane of the magnet. The thermal noise is added in three directions ($\hat x$, $\hat y$ and $\hat z$) with zero mean and $\langle H_{\mathrm{noise}}^2 \rangle  = 2 \alpha k_BT / [(\gamma M_s \mathrm{Vol.})]$ in units [$\rm Oe^2/s$] \cite{li2004thermally}.

\subsection{Device operation}

The p-bit shown in FIG.~\ref{fi:fig5}c is a series-resistance controlled device where the transistor resistance can be made much smaller or much larger compared to the fluctuating MTJ resistance. Therefore, the operation of the p-bit does not require manipulating the magnetization of the free layer unlike in standard spin-transfer-torque MRAM cells. However, the current flowing through the fixed layer of the MTJ produces a spin-polarized spin current that can unintentionally torque the magnet.  We assume that this current is given as $\vec{I}_s = P I_{\mathrm{MTJ}} \hat z $, where $\hat z$ is the fixed layer orientation and $P$ is an interface polarization that can be related to TMR \cite{datta2012voltage}. This spin-current is fed back to the sLLG solver and fully accounted for in the calculation of magnetization in our simulations, however for the circular LBM with a large demagnetization field used here, its effects are negligible \cite{faria2017low}. Using these models, FIG.~\ref{fi:fig5}c shows transient SPICE simulations of a single p-bit output, $V_{\rm OUT}$ for 1000 samples where $V_{\rm IN}$ is rapidly swept in 2 ns. The range of stochastic outputs is bounded by a distribution of resistances ranging from $R_P$ to $R_{AP}$. The ensemble average shows an approximate hyperbolic tangent behavior that allows the mapping shown in Eq.~\ref{eq:mtj_pbit}.
 
The inset of FIG.~\ref{fi:fig5}d shows the autocorrelation time of the circular in-plane magnet with a lifetime of  $\approx 100$~ps. The fluctuations for a circular magnet is expected to be faster compared to a magnet with perpendicular anisotropy due to the strong demagnetizing field that keeps the  magnetization vector in the easy plane of the magnet \cite{lopez2002transition}. The very short lifetime of such a circular low barrier magnet could allow very fast and efficient sampling times, as long as the interconnection network operates faster than these timescales. In present simulations, the resistive network operates instantaneously with an ideal operational amplifier therefore this requirement is met naturally, however in real implementations the synapse needs to be designed carefully.

The second requirement, the need for sequential updating of each p-bit is met naturally since the probability of simultaneous flips among p-bits is extremely unlikely, therefore hardware p-bits evolve autonomously without a synchronizing clock, effectively going through a random update order that does not affect their final distribution.


\subsection{Stochastic MRAM-based p-bit vs Quantum Boltzmann Law}
In FIG.~\ref{fi:fig6}, using full SPICE simulations for a 40 p-bit network we compute the joint probability density of a $M=4$ spin ferromagnetic chain ($J_{i,j}=+1$) using 10 replicas, with $\beta=0.5$ and $\Gamma_x=10$. Unlike FIG.~\ref{fi:fig2}, no symmetry breaking field is applied and the network is asynchronously operated for $t_{sim}=250$ ns, with a time step of 1 ps.  All analog voltage values at the end of the SPICE simulation are thresholded ($> 0 \rm\ V \equiv 1$, $< 0 \rm \ V \equiv -1$) and a time-average is obtained similar to the PSL averaging after converting the state of each p-bit to binary and then to decimal. The results from the full device models seem to be in good agreement with the exact solution obtained from Eq.~\ref{eq:qboltzmann} and the behavioral PSL equations that are included for reference. Note the suppression of states $5=(0101)_2$ and $10=(1010)_2$ that correspond to the energetically unfavorable antiferromagnetic configurations $\ket{\downarrow \uparrow \downarrow \uparrow}$ and $\ket{\uparrow \downarrow \uparrow \downarrow}$, respectively. The agreement between the full SPICE models with the behavioral and exact solutions establishes the feasibility of the proposed quantum circuit emulator.

\subsection{Projected performance improvement}
In annealing algorithms, a key parameter is the time-to-solution (TTS) that is defined as the total time a solver requires to reach the desired
answer of a problem with a predefined accuracy \cite{albash2018demo}. TTS clearly depends on the intrinsic hardware substrate that is used to implement the algorithm  but also on the type of the problem and the required accuracy in the solution. Since the type of problem can have varied scaling properties with no generic answers \cite{albash2018demo,boixo_evidence_2014,montenegro2006mathematical}, here we attempt to define a basic hardware unit, which is the time to provide a spin-flip attempt as defined by Eq.~\ref{eq:pbit}. As shown in the inset of Fig.~\ref{fi:fig5}, the correlation time of an in-plane circular magnet can be as low as about 100 ps due to a precessional fluctuation mechanism found in such low barrier magnets \cite{hassan2019low}. Assuming a nearest  neighbor 3D classical network that is mapped from a 2D quantum network, we assume that the synapse delay due to a crossbar structure can be much faster than magnetic fluctuations, ensuring all spin flip updates use up-to-date information and are useful. In such a scenario having $N = 10^6$ spins that are operating autonomously  with a 0.1 to 1 ns correlation times, we project that the spin-flip rate can be 0.1 to 1 petaflips per second, which is orders of magnitude faster than present day CPU implementations \cite{boixo_evidence_2014} as well as parallelized GPU implementations \cite{fang2014parallel}. A detailed projection with power estimates can be found in \cite{sutton2019autonomous}.  Finally, we note that such ~GHz rate fluctuations of low barrier magnets have been not only theoretically predicted \cite{hassan2019low} but also experimentally observed in in-plane magnets \cite{pufall2004large}. 

\section{Conclusion}
We have presented a scalable, room-temperature quantum emulator using stochastic p-bits that can be built by a simple modification of the existing 1T/1MTJ cell of the eMRAM technology.  The proposed emulator uses physical replicas for repeated Trotter slices used in software Quantum Monte Carlo methods. Having physical replicas for each slice could enable better scaling properties for quantum annealing compared to classical annealing as discussed in \cite{santoro2002theory}, since choosing the optimal number of replicas or probing each replica separately to find better energy minima is possible in a physically engineered design, unlike in real quantum systems \cite{heim2015quantum}. The electrical control of annealing parameters, inverse temperature ($\beta$) and transverse field ($\Gamma_x$), could allow a very large number of q-bits to be reliably emulated with room temperature p-bits. Using  conventional electronic devices such as GPU's or FPGA's to implement the synapses, it should be possible to engineer complicated interactions that extend beyond nearest neighbors and/or involve $k$-body interactions ($k>2$). We note that even though the  ``sign problem'' limits the universal use of our p-computer, a  large number of practically relevant quantum systems could be efficiently emulated by it, considering a large number of optimization problems have been mapped on to the Transverse Ising Hamiltonian \cite{lucas2014ising}. Our results provide a method of emulating quantum systems with probabilistic hardware in advance of a  scalable universal quantum computer.

\begin{acknowledgments}
KYC is grateful to Brian M. Sutton for helpful discussions. This work was supported in part by ASCENT, one of six centers in JUMP, a Semiconductor Research Corporation (SRC) program sponsored by DARPA.  
\end{acknowledgments}

\section*{Appendix A : Mapping Quantum Heisenberg Model to a Classical System}
The Heisenberg Hamiltonian emulated in Section \ref{sec:heis} is,
\begin{equation}
\mathcal{H}_Q \hspace{-4pt} = -\hspace{-1pt} \hspace{-1pt}\left(\hspace{-1pt}\hspace{-2pt} \sum_{i}^{M} J_z \sigma^z_i \sigma^z_{i+1} \hspace{-3pt}+ \hspace{-3pt} J_x \sigma^x_i \sigma^x_{i+1}  \hspace{-3pt}+ \hspace{-3pt}J_y \sigma^y_i \sigma^y_{i+1} \hspace{-3.5pt}+ \hspace{-3.5pt} \Gamma_x \sigma^x_i \hspace{-4pt}\right) \hspace{-3pt}\end{equation}
Following \cite{suzuki1976relationship,barma1978classical}, we divide this Hamiltonian into three non-commuting parts, i.e., $\mathcal{H}_Q=\mathcal{H}_0+\mathcal{H}_1+\mathcal{H}_2$, where 
\begin{eqnarray}
\mathcal{H}_0 &=& \hspace{-3.5pt}- \hspace{-5.5pt}\sum_{i=1}^{M}{J_{z}\sigma_{i}^z\sigma^z_{i+1}}\\
\mathcal{H}_1 &=&\hspace{-3.5pt} - \hspace{-5.5pt}\sum_{i=1,3,\cdots}^{M}\hspace{-6pt}{\left(J_{x}\sigma_{i}^x\sigma^x_{i+1}+J_{y}\sigma_{i}^y\sigma^y_{i+1}\right)}-\hspace{-3.5pt}\cfrac{1}{2}\sum_{i}^{M}{\Gamma_x\sigma_i^x}\\
\mathcal{H}_2 &=&\hspace{-3.5pt} - \hspace{-5.5pt}\sum_{i=2,4,\cdots}^{M}\hspace{-6pt}{\left(J_{x}\sigma_{i}^x\sigma^x_{i+1}+J_{y}\sigma_{i}^y\sigma^y_{i+1}\right)}-\hspace{-3.5pt}\cfrac{1}{2}\sum_{i}^{M}{\Gamma_x\sigma_i^x}
\end{eqnarray}

The $n$-th approximant of the Suzuki-Trotter transformation for this Hamiltonian is then given by
\begin{equation}
\begin{split}
Z_Q^{(n)}= \sum_{\alpha_1,\alpha_2,\cdots,\alpha_{2n}}&{\Big[\prod_{k=1}^{2n}{e^{-\beta\mathcal{H}_0\left(\alpha_k\right)/2n}}}\nonumber\\
&\times\prod_{k=1,3,\cdots}^{2n-1}\langle \alpha_k\rvert e^{-\beta\mathcal{H}_1/n}\lvert\alpha_{k+1}\rangle\\
&\times\prod_{k=2,4,\cdots}^{2n}\langle \alpha_k\rvert e^{-\beta\mathcal{H}_2/n}\lvert\alpha_{k+1}\rangle\Big] \nonumber
\end{split}
\end{equation}
and the classical system becomes,
\begin{equation}
\begin{split}
\mathcal{H}_{d+1}&=\sum_{k=1,2,3,\cdots}^{2n}{\cfrac{1}{2n}\mathcal{H}_0\left(\alpha_k\right)}\\
&-\cfrac{1}{\beta}\sum_{k=1,3,\cdots}^{2n-1}{\ln\langle\alpha_k\rvert e^{-\beta\mathcal{H}_1/n}\lvert\alpha_{k+1}\rangle}\\
&-\cfrac{1}{\beta}\sum_{k=2,4,\cdots}^{2n}{\ln\langle\alpha_k\rvert e^{-\beta\mathcal{H}_2/n}\lvert\alpha_{k+1}\rangle}
\end{split}
\label{eq:Hclassical}
\end{equation}
with periodicity along ($d+1)^{\rm th}$ dimension such that $\lvert\alpha_{2n+1}\rangle=\lvert\alpha_{1}\rangle$. Also notice that any $k$-th replica can also be written more explicitly using Dirac's bra-ket notation in terms of the constituent spins of that replica as $\lvert\alpha_{k}\rangle=\lvert m_{1,k}m_{2,k}\cdots m_{M,k}\rangle$ which is actually a $2^M\times1$ column vector and $m_{i,j}$ denotes $i$-th spin of $j$-th replica.

Then the first summation on the right hand side of Eq. (\ref{eq:Hclassical}) can be written as
\begin{equation}
\sum_{k=1,2,3,\cdots}^{2n}{\cfrac{1}{2n}\,\mathcal{H}_0\left(\alpha_k\right)}=\sum_{k=1}^{2n}\sum_{i=1}^{M}{\cfrac{J_z}{2n}\,m_{i,k}m_{i+1,k}+\cfrac{\Gamma_z}{2n}\,m_{i,k}}
\end{equation} 

In order to  evaluate $\ln\langle\alpha_k\rvert e^{-\beta\mathcal{H}_1/n}\lvert\alpha_{k+1}\rangle$ and $\ln\langle\alpha_k\rvert e^{-\beta\mathcal{H}_2/n}\lvert\alpha_{k+1}\rangle$,  we start by repeatedly using the following identity of the Kronecker product:
\begin{equation}
e^{\mathbf{A}\otimes \mathbf{I}_B+\mathbf{I}_A\otimes \mathbf{B}} = e^{\mathbf{A}}\otimes e^{\mathbf{B}} 
\end{equation}
to write the following:
\begin{equation}
e^{-\beta\mathcal{H}_1/n} = e^{-\frac{\beta}{n}\zeta}\otimes e^{-\frac{\beta}{n}\zeta}\otimes\cdots\otimes e^{-\frac{\beta}{n}\zeta}
\label{eq:resIdentity1}
\end{equation}
where 
\begin{equation}
\begin{split}
\mathcal{\zeta} =& - J_x\left(\sigma^x\otimes\mathbf{I}_2\right)\left(\mathbf{I}_2\otimes\sigma^x\right)-J_y\left(\sigma^y\otimes\mathbf{I}_2\right)\left(\mathbf{I}_2\otimes\sigma^y\right)\\
&-\cfrac{1}{2}\,\Gamma_x\left(\sigma^x\otimes\mathbf{I}_2+\mathbf{I}_2\otimes\sigma^x\right)
\end{split}
\end{equation}
and is a $4\times 4$ matrix representing a two-body Hamiltonian.

Here we note that $\lvert\alpha_{k}\rangle$ can also be partitioned in terms of Kronecker products of two spin systems as
\begin{equation}
\lvert\alpha_{k}\rangle = \lvert m_{1,k}m_{2,k}\rangle \otimes \lvert m_{3,k}m_{4,k}\rangle\otimes\cdots\otimes\lvert m_{M-1,k}m_{M,k}\rangle.
\end{equation}

With the definition of $\lvert\alpha_{k}\rangle$ above in mind, we also repeatedly use another Kronecker product identity:
\begin{equation}
\left(\mathbf{A}\otimes\mathbf{B}\right)\left(\mathbf{C}\otimes\mathbf{D}\right) = \left(\mathbf{AC}\right)\otimes\left(\mathbf{BD}\right)
\end{equation}
to write 
\begin{equation}
\begin{split}
&\langle\alpha_k\rvert e^{-\beta\mathcal{H}_1/n}\lvert\alpha_{k+1}\rangle\\&=\prod_{i=1,3,\cdots}^{M}{\langle m_{i,k}m_{i+1,k}\rvert e^{-\beta\mathcal{\zeta}/n}\lvert m_{i,k+1}m_{i+1,k+1}\rangle}
\end{split}
\end{equation}
where we have also made use of Eq. (\ref{eq:resIdentity1}). Taking e-base logarithm on both sides, we finally get
\begin{equation}
\begin{split}
&\ln\langle\alpha_k\rvert e^{-\beta\mathcal{H}_1/n}\lvert\alpha_{k+1}\rangle\\&=\sum_{i=1,3,\cdots}^{M}{\ln\langle m_{i,k}m_{i+1,k}\rvert e^{-\beta\mathcal{\zeta}/n}\lvert m_{i,k+1}m_{i+1,k+1}\rangle}.
\end{split}
\end{equation}
In a similar manner, we can also write,
\begin{equation}
\begin{split}
&\ln\langle\alpha_k\rvert e^{-\beta\mathcal{H}_2/m}\lvert\alpha_{k+1}\rangle\\&=\sum_{i=2,4,\cdots}^{M}{\ln\langle m_{i,k}m_{i+1,k}\rvert e^{-\beta\zeta/n}\lvert m_{i,k+1}m_{i+1,k+1}\rangle}.
\end{split}
\end{equation}

Next, we evaluate the 4$\times$4 density matrix:
\begin{equation}
e^{-\beta {\varsigma}/n}=\left[\begin{array}{cccc}
X_1 & X_5 & X_5 & X_2\\
X_5 & X_3 & X_4 & X_5\\
X_5 & X_4 & X_3 & X_5\\
X_2 & X_5 & X_5 & X_1
\end{array}\right]
\end{equation}

The corresponding $X_i$ are given by: 
{\footnotesize{
\begin{equation*}
X=\sqrt{{{\Gamma_x}^{2}}+{ {{J}_{y}}^{2}}} 
\end{equation*}
\begin{equation*}
{{X}_{1}}=\frac{1}{2}{{e}^{\frac{\beta {{J}_{x}}}{n}}}\hspace{-4pt}\left[ \cosh \left(\hspace{-2pt}\frac{\beta }{n}X\hspace{-2pt} \right)\hspace{-2pt}-\hspace{-2pt}\left(\hspace{-2pt}\frac{{{J}_{y}}}{X} \hspace{-2pt}\right)\hspace{-2pt}\sinh \hspace{-2pt}\left(\hspace{-2pt}\frac{\beta }{n}X \hspace{-2pt}\right) \right]+\frac{1}{2}{{e}^{\frac{-\beta \left( {{J}_{x}}\hspace{-1pt}-\hspace{-1pt}{{J}_{y}}\hspace{-1pt} \right)}{n}}}
\end{equation*}
\begin{equation*}
{{X}_{2}}=\frac{1}{2}{{e}^{\frac{\beta {{J}_{x}}}{n}}}\hspace{-4pt}\left[ \cosh \left(\hspace{-2pt}\frac{\beta }{n}X\hspace{-2pt} \right)\hspace{-2pt}-\hspace{-2pt}\left(\hspace{-2pt}\frac{{{J}_{y}}}{X} \hspace{-2pt}\right)\hspace{-2pt}\sinh \hspace{-2pt}\left(\hspace{-2pt}\frac{\beta }{n}X \hspace{-2pt}\right) \right]-\frac{1}{2}{{e}^{\frac{-\beta \left( {{J}_{x}}\hspace{-1pt}-\hspace{-1pt}{{J}_{y}}\hspace{-1pt} \right)}{n}}}
\end{equation*}
\begin{equation*}
{{X}_{3}}=\frac{1}{2}{{e}^{\frac{\beta {{J}_{x}}}{n}}}\hspace{-4pt}\left[ \cosh \left(\hspace{-2pt}\frac{\beta }{n}X\hspace{-2pt} \right)\hspace{-2pt}+\hspace{-2pt}\left(\hspace{-2pt}\frac{{{J}_{y}}}{X} \hspace{-2pt}\right)\hspace{-2pt}\sinh \hspace{-2pt}\left(\hspace{-2pt}\frac{\beta }{n}X \hspace{-2pt}\right) \right]+\frac{1}{2}{{e}^{\frac{-\beta \left( {{J}_{x}}\hspace{-1pt}+\hspace{-1pt}{{J}_{y}}\hspace{-1pt} \right)}{n}}}
\end{equation*}
\begin{equation*}
{{X}_{4}}=\frac{1}{2}{{e}^{\frac{\beta {{J}_{x}}}{n}}}\hspace{-4pt}\left[ \cosh \left(\hspace{-2pt}\frac{\beta }{n}X\hspace{-2pt} \right)\hspace{-2pt}+\hspace{-2pt}\left(\hspace{-2pt}\frac{{{J}_{y}}}{X} \hspace{-2pt}\right)\hspace{-2pt}\sinh \hspace{-2pt}\left(\hspace{-2pt}\frac{\beta }{n}X \hspace{-2pt}\right) \right]-\frac{1}{2}{{e}^{\frac{-\beta \left( {{J}_{x}}\hspace{-1pt}+\hspace{-1pt}{{J}_{y}}\hspace{-1pt}\right)}{n}}}
\end{equation*}
\begin{equation*}
{{X}_{5}}=\frac{1}{2}{{e}^{\frac{\beta {{J}_{x}}}{n}}}\frac{\Gamma_x }{X}\sinh \left( \frac{\beta }{n}X \right).	
\end{equation*}}}

We then use the following energy relation (a justification of the energy model will be presented in Appendix B):
\begin{eqnarray}
&-&\cfrac{1}{\beta}\ln \langle m_{i,k}m_{i+1,k}\rvert e^{-\beta\mathcal{\varsigma}/n}\lvert m_{i,k+1}m_{i+1,k+1}\rangle \nonumber \\ 
&=&2\epsilon -{{t}_{1}}\left( {{m}_{i,k}}{{m}_{i,k+1}}+{{m}_{i+1,k}}{{m}_{i+1,k+1}} \right)\nonumber \\ 
&-&{{t}_{2}}\left( {{m}_{i,k}}{{m}_{i+1,k+1}}+{{m}_{i+1,k}}{{m}_{i,k+1}} \right)\nonumber \\ 
&-&{{t}_{3}}\left( {{m}_{i,k}}{{m}_{i+1,k}}+{{m}_{i,k+1}}{{m}_{i+1,k+1}} \right)\nonumber \\
&-&{{t}_{4}}{{m}_{i,k}}{{m}_{i,k+1}}{{m}_{i+1,k}}{{m}_{i+1,k+1}}
\label{eq:enMod}
\end{eqnarray}
where $\epsilon$ is a constant that we ignore and 
\begin{eqnarray*}
	t_0 &=&\cfrac{1}{2n} J_{z} \\ 
	t_1 &=& \cfrac{1}{8\beta}\left(\ln X_1 - \ln X_2 +\ln X_3 -\ln X_4\right)  \label{eq:t1xyz}\\
	t_2 &=& \cfrac{1}{8\beta}\left(\ln X_1 - \ln X_2 -\ln X_3 +\ln X_4\right)  \label{eq:t2xyz}\\
	t_3 &=& \cfrac{1}{8\beta}\left(\ln X_1 + \ln X_2 -\ln X_3 -\ln X_4\right) \label{eq:t3xyz}\\
	t_4 &=& \cfrac{1}{8\beta}\left(\ln X_1 + \ln X_2 +\ln X_3 +\ln X_4 - 4\ln X_5\right)\label{eq:t4xyz}
\end{eqnarray*}
This corresponds to the energy model for the Heisenberg Hamiltonian as shown in Fig.~\ref{fi:fig3}.
\section*{Appendix B : Justification of using form of the energy model in Appendix A}
\label{app:appB}
We start by simplifying the notation such that $m_{i,k}\equiv m_1$, $m_{i+1,k}\equiv m_2$, $m_{i,k+1}\equiv m_3$, and $m_{i+1,k+1}\equiv m_4$ and put different $I_1$  values for different configurations of $\{m_2,m_3,m_4\}$ into a truth table as shown in Table \ref{tab:tab1} with following definitions:
\begin{eqnarray}
f_1&=&\cfrac{1}{2}\ln{\left(\cfrac{X_1}{X_5}\right)}\\
f_2&=&\cfrac{1}{2}\ln{\left(\cfrac{X_5}{X_2}\right)}\\
f_3&=&\cfrac{1}{2}\ln{\left(\cfrac{X_5}{X_4}\right)}\\
f_4&=&\cfrac{1}{2}\ln{\left(\cfrac{X_3}{X_5}\right)}
\end{eqnarray}

\begin{table}[!ht]
	\centering
	\caption{Truth table for $I_1$.}
	\vspace{8pt}
	\begin{tabular}{@{}ccc|c@{}}
		$s_3$&$s_4$&$s_2$&\multicolumn{1}{c}{ $I_1=\cfrac{1}{2}\ln\left(\cfrac{P\left(m_1=+1|m_3,m_4,m_2\right)}{P\left(m_1=-1|m_3,m_4,m_2\right)}\right)$} \\ [0.75em]\hline
		1&1&1&$f_1$\\
		1&1&0&$f_2$\\
		1&0&1&$f_3$\\
		1&0&0&$f_4$\\
		0&1&1&$-f_4$\\
		0&1&0&$-f_3$\\
		0&0&1&$-f_2$\\
		0&0&0&$-f_1$
	\end{tabular}
	\label{tab:tab1}
\end{table}
We have also used the notation that $s_i\in\{0,1\}, (i\in\{1,2,3,4\})$ is the binary counterpart of the bipolar spin $m_i\in\{-1,1\}$.

In the binary representation, we can cast $I_1$ into the following form:
\begin{equation}
\begin{split}
I_1 &= f_1s_3s_4s_2-f_1\bar{s}_3\bar{s}_4\bar{s}_2+f_2s_3s_4\bar{s}_2-f_2\bar{s}_3\bar{s}_4s_2\\
&+f_3s_3\bar{s}_4s_2-f_3\bar{s}_3{s}_4\bar{s}_2+f_4s_3\bar{s}_4\bar{s}_2-f_4\bar{s}_3{s}_4{s}_2
\end{split}
\label{eq:eqI1_bin}
\end{equation}
We use the following two transformations to switch from $s_i$ to $m_i$:
\begin{eqnarray}
m_i&\to&\cfrac{1+s_i}{2}\\
\bar{m}_i&\to&\cfrac{1-s_i}{2},
\end{eqnarray}  
we can recast Eq.(\ref{eq:eqI1_bin}) into its bipolar form as
{\footnotesize{
\begin{equation}
\begin{split}
I_1 &= \cfrac{f_1}{8}\left[(1+m_3)(1+m_4)(1+m_2)-(1-m_3)(1-m_4)(1-m_2)\right]\\
&+\cfrac{f_2}{8}\left[(1+m_3)(1+m_4)(1-m_2)-(1-m_3)(1-m_4)(1+m_2)\right]\\
&+\cfrac{f_3}{8}\left[(1+m_3)(1-m_4)(1+m_2)-(1-m_3)(1+m_4)(1-m_2)\right]\\
&+\cfrac{f_4}{8}\left[(1+m_3)(1-m_4)(1-m_2)(1-m_3)(1+m_4)(1+m_2)\right]
\end{split}
\label{eq:eqI1_bip}
\end{equation}}}

Upon simplification and re-arrangement of  the terms we get,
\begin{equation}
\begin{split}
I_1 &= \left(\cfrac{f_1}{4}-\cfrac{f_2}{4}+\cfrac{f_3}{4}-\cfrac{f_4}{4}\right)m_2\\&+\left(\cfrac{f_1}{4}+\cfrac{f_2}{4}+\cfrac{f_3}{4}+\cfrac{f_4}{4}\right)m_3\\
&+\left(\cfrac{f_1}{4}+\cfrac{f_2}{4}-\cfrac{f_3}{4}-\cfrac{f_4}{4}\right)m_4\\&+\left(\cfrac{f_1}{4}-\cfrac{f_2}{4}-\cfrac{f_3}{4}+\cfrac{f_4}{4}\right)m_2m_3m_4
\end{split}
\label{eq:eqI1_bip3}
\end{equation}

Integrating Eq.(\ref{eq:eqI1_bip3}) with respect to $m_1$ and multiplying by $\left(-\cfrac{1}{\beta}\right)\,$, we partially get the energy model
\begin{equation}
\begin{split}
E &= -\cfrac{1}{\beta}\left(\cfrac{f_1}{4}-\cfrac{f_2}{4}+\cfrac{f_3}{4}-\cfrac{f_4}{4}\right)m_1m_2\\&-\cfrac{1}{\beta}\left(\cfrac{f_1}{4}+\cfrac{f_2}{4}+\cfrac{f_3}{4}+\cfrac{f_4}{4}\right)m_1m_3\\
&-\cfrac{1}{\beta}\left(\cfrac{f_1}{4}+\cfrac{f_2}{4}-\cfrac{f_3}{4}-\cfrac{f_4}{4}\right)m_1m_4\\&-\cfrac{1}{\beta}\left(\cfrac{f_1}{4}-\cfrac{f_2}{4}-\cfrac{f_3}{4}+\cfrac{f_4}{4}\right)m_1m_2m_3m_4\\&+K_1\left(m_2,m_3,m_4\right)\\
\end{split}
\label{eq:eqE}
\end{equation}
where $K_1$ is a function of $m_2$, $m_3$ and $m_4$ but independent of $m_1$.

Defining,
\begin{eqnarray}
t_3 &=& \cfrac{1}{4\beta}\left(f_1-f_2+f_3-f_4\right)\nonumber\\
&=&\cfrac{1}{8\beta}\left(\ln X_1 + \ln X_2 -\ln X_3 -\ln X_4\right) \label{eq:t3xyz}\\
t_1 &=& \cfrac{1}{4\beta}\left(f_1+f_2+f_3+f_4\right)\nonumber\\
&=&\cfrac{1}{8\beta}\left(\ln X_1 - \ln X_2 +\ln X_3 -\ln X_4\right)  \label{eq:t1xyz}\\
t_2 &=& \cfrac{1}{4\beta}\left(f_1+f_2-f_3-f_4\right)\nonumber\\
&=&\cfrac{1}{8\beta}\left(\ln X_1 - \ln X_2 -\ln X_3 +\ln X_4\right)  \label{eq:t2xyz}\\
t_4 &=& \cfrac{1}{4\beta}\left(f_1-f_2-f_3+f_4\right) \nonumber\\
&=&\cfrac{1}{8\beta}\left(\ln X_1 + \ln X_2 +\ln X_3 +\ln X_4 - 4\ln X_5\right)  \label{eq:t4xyz}
\end{eqnarray}
we get the simplified energy expression:
\begin{equation}
\begin{split}
&E\left(m_1,m_2,m_3,m_4\right)\Big|_{m_1}=-t_1m_1m_3-t_2m_1m_4\\&-t_3m_1m_2-t_4m_1m_2m_3m_4+K_1\left(m_2,m_3,m_4\right)
\end{split}
\label{eq:K1}
\end{equation} 

Repeating the whole procedure for $I_2$, $I_3$ and $I_4$ separately, give us
\begin{eqnarray}
K_1\left(m_2,m_3,m_4\right)&=&-t_1m_2m_4-t_2m_2m_3 \nonumber\\
&&+K_2\left(m_3,m_4\right)\\
K_2\left(m_3,m_4\right)&=&-t_3m_3m_4+K_3\left(m_4\right)\\
K_3\left(m_4\right)&=&0.
\label{eq:K3}
\end{eqnarray}

Putting Eqs.(\ref{eq:K1}-\ref{eq:K3}) together, gives us the energy model used in Eq.(\ref{eq:enMod}).

\end{document}